\documentclass[draft]{agujournal2019}
\usepackage{url} 
\usepackage{lineno}
\draftfalse

\journalname{JGR: Solid Earth}

\begin{document}

%
%


\title{An eight-year-long low-frequency earthquake catalog for Southern Cascadia}

%
%




\authors{Ariane Ducellier\affil{1}, Kenneth C. Creager\affil{1}}


\affiliation{1}{University of Washington}




\correspondingauthor{Ariane Ducellier}{ducela@uw.edu}




\begin{keypoints}
\item We establish an eight-year-long low-frequency earthquake catalog for southern Cascadia.
\item Families from the subduction zone are mainly active during the big tremor episodes, with down dip families more active than up dip families.
\item LFE activity is sensitive to tidal stress changes.
\end{keypoints}

%
%


\begin{abstract}
Low-frequency earthquakes (LFEs) are small magnitude earthquakes, with typical magnitude less than 2, and reduced amplitudes at frequencies greater than 10 Hz relative to  ordinary small earthquakes. Their occurrence is often associated with tectonic tremor and slow-slip events along the plate boundary in subduction zones and occasionally transform fault zones. They are usually grouped into families, with all the earthquakes of a given family originating from the same small patch on the plate interface and recurring more or less episodically in a bursty manner. In this study, we extend the LFE catalog obtained by ~\citeA{PLO_2015} during an episode of high tremor activity in  April 2008, to the 8-year period 2004-2011. All of the tremor in the ~\citeA{BOY_2015} catalog south of 42 degrees North has associated LFE activity, but we have identified several smaller episodes of LFEs without associated tremor activity, and extend their catalog forward and backward by a total of about 3 years. As in northern Cascadia, the down-dip LFE families have recurrence intervals several times smaller than the up-dip families. For the April 2008 Episodic Tremor and Slip event, the best recorded LFE families exhibit a strong tidal Coulomb stress sensitivity starting 1.5 days after the rupture front passes by each LFE family.  The southernmost LFE family, which has been interpreted to be on the subduction plate boundary, near the up-dip limit of tremor, has a very short recurrence time. Also, these LFEs tend to occur during times when predicted tidal Coulomb stress is discouraging slip on the plate boundary. Both observations suggest this LFE family may be on a different fault.
\end{abstract}

\section*{Plain language summary}

Low-frequency earthquakes (LFEs) are small magnitude earthquakes, with typical magnitude less than 2, and reduced amplitudes at frequencies greater than 10 Hz relative to ordinary small earthquakes. They are usually grouped into families, with all the earthquakes of a given family originating from the same small patch on the plate interface. LFEs tend to happen in bursts, with dozen of earthquakes occurring within a few days, followed by weeks or months of quiet. In this study, we extend the LFE catalog obtained by ~\citeA{PLO_2015} to the 8-year period 2004-2011. As in northern Cascadia, the LFE families located at greater depth (down dip) are more active, with shorter time intervals between two earthquake bursts, than the LFE families located at shallower depth (up dip). Earth tides also have an influence on LFE activity: LFEs are more likely to occur when the stress changes associated with the tides favor shear failure on the plate boundary.

%
%

%


%
%
%
%

\section{Introduction}

Tectonic tremor is a weak but persistent shaking of the Earth that has been discovered in many subduction zones and some strike-slip faults throughout the world ~\cite{BER_2011}. Tremor is observed on seismograms as apparent noise whose amplitude is modulated in time in a similar manner at stations that are dozens of kilometers apart from each other ~\cite{OBA_2002}. It is characterized by a long (several seconds to many minutes), low amplitude seismic signal, emergent onsets, and an absence of clear impulsive phases. Tremor can be explained as a swarm of low-frequency earthquakes (LFEs) ~\cite{SHE_2007_nature}, that is small magnitude earthquakes (M $\sim$ 1) whose dominant frequency is clearly low (1-10 Hz) compared with ordinary tiny earthquakes (up to 20 Hz). The source of the tremor and the LFEs is located on the plate boundary ~\cite{SHE_2006,BOS_2012,AUD_2016}, and their focal mechanisms represent shear slip on a low-angle thrust fault dipping in the same direction as the plate interface ~\cite{IDE_2007_GRL,BOS_2012,ROY_2014}. LFEs are usually grouped into families of earthquakes, with all the earthquakes of a given family originating from the same small patch on the plate interface, and recurring more or less episodically in a bursty manner. Dozens of earthquakes are thus recorded within a few hours or days during a burst of LFE activity, followed by weeks or months of quiet, with very few earthquakes. In subduction zones such as Nankai and Cascadia, tectonic tremor and LFE observations are spatially and temporally correlated with slow-slip observations ~\cite{ROG_2003,OBA_2004}. Due to this correlation, these paired phenomena have been called Episodic Tremor and Slip (ETS). \\

The relatively short recurrence of slow-slip and tremor episodes results in a rich history both in space and time and reveals potential patterns.  These tremor activity histories have allowed scientists to see complete cycles, which is typically not possible to explore in traditional earthquake catalogs. However, most of the work on LFEs has been focused on detecting LFEs during periods of high tremor activity, grouping them into families of earthquakes, and locating the source of the LFE families. Longer catalogs (several years) have been established for LFE families in Mexico (two-year long catalog by ~\citeA{FRA_2014}), the San Andreas Fault (fifteen-year-long catalog by ~\citeA{SHE_2017}), Washington State (five-year-long catalog by ~\citeA{SWE_2019} and two-year-long catalog by ~\citeA{CHE_2017_JGR, CHE_2017_G3}), Vancouver Island (ten-year-long catalog by ~\citeA{BOS_2015}), New Zealand (eight-year-long catalog by ~\citeA{BAR_2018}), and Japan (twelve-year-long catalog by ~\citeA{NAK_2017}, eight-year-long catalog by ~\citeA{OHT_2017}, eleven-year-long catalog by ~\citeA{KAT_2020}, and Japan Meteorological Agency (JMA) catalog since 1999 ~\cite{KAT_2003}). These studies have shown that the recurrence behavior of LFE families varies a lot between seismic regions, and inside the same seismic region. In northern Washington, ~\citeA{SWE_2019} have identified and characterized four different LFE families that span the width of the transition zone in the Cascadia Subduction Zone beneath western Washington State. They found that the LFEs swarm duration, recurrence interval, and cluster size decrease systematically with increasing depth. On the San Andreas Fault, ~\citeA{SHE_2017} observed a large diversity of recurrence behaviors among the LFE families, from semicontinuous to highly episodic. Particularly, two families exhibited bimodal recurrence patterns (about 3 and 6 days for the first one, and about 2 and 4 days for the second one). Moreover, he observed an increase in the LFE rate after the 2004 Parkfield earthquake. \\

~\citeA{PLO_2015} have detected LFEs in southern Cascadia during the April 2008 ETS event using seismic data from the EarthScope Flexible Array Mendocino Experiment (FAME). They used a combination of autodetection methods and visual identification to obtain the initial templates. Then, they recovered higher signal-to-noise LFE signals using iterative network cross-correlation. They found that the LFE families on the southern Cascadia Subduction Zone were located above the plate boundary model of ~\citeA{MCC_2006}, with a large distribution of depths (28-47 km). However, they suggested that this may be because the plate boundary model is biased deep. Three additional LFE families were found on two strike-slip faults, the Maacama and Bucknell Creek faults, which are part of the San Andreas Fault zone. \\

When the hard work of detecting LFEs and identifying LFE families has been carried out, and enough (a few hundred) LFEs have been identified for a given family, a template waveform can be obtained by stacking all the waveforms corresponding to all the LFEs identified. Once a template is available, additional LFEs can be found by cross-correlating seismic data with the template, and assuming that an LFE is occurring whenever the value of the cross-correlation is higher than a chosen threshold.  The signal-to-noise ratios are low, so LFEs can be best identified by stacking the cross-correlation functions of multiple stations. In this study, we first use the catalog established by ~\citeA{PLO_2015} for the months of March and April 2008 to create templates for the temporary seismic stations of the FAME experiment. We then use these templates to extend the catalog to the whole period when the FAME experiment was running, between July 2007 and June 2009. Next, we use the LFE detections from the 2007-2009 period to create templates for the permanent stations of three seismic networks in northern California. These new templates allow us to extend the LFE catalog to the period 2004-2011.

\section{Data}

We used both seismic data from the temporary EarthScope Flexible Array Mendocino Experiment (FAME) distributed by Incorporated Research Institutions for Seismology (IRIS), and seismic data from three permanent seismic networks distributed by the Northern California Earthquake Data Center (NCEDC). The FAME network was installed in northern California between July 2007 and June 2009. The three permanent networks are Berkeley Digital Seismic Network (BK), Northern California Seismic Network (NC), and Plate Boundary Observatory Strain and Seismic Data (PB). We used both one-component and three-components seismic stations. Depending on availability, we used channels BHZ, EHZ, HHZ, or SHZ as we are mainly interested in the frequency band 1-10Hz. We restricted ourselves to seismic stations less than 100 kilometers away from the epicenter of an LFE family, as we do not expect to have good signal-to-noise ratio for stations located farther away. The complete list of seismic stations and channels used in this study is given in the Supplementary Material (Tables S1 and S2). Figure 1 shows a map of the locations of the LFE families, and of the locations of the seismic stations. We can see that we have a good coverage of the area, and most LFE families are surrounded by several seismic stations. \\

\begin{figure}
\noindent\includegraphics[width=\textwidth, trim={0.5cm 7cm 1.5cm 7cm},clip]{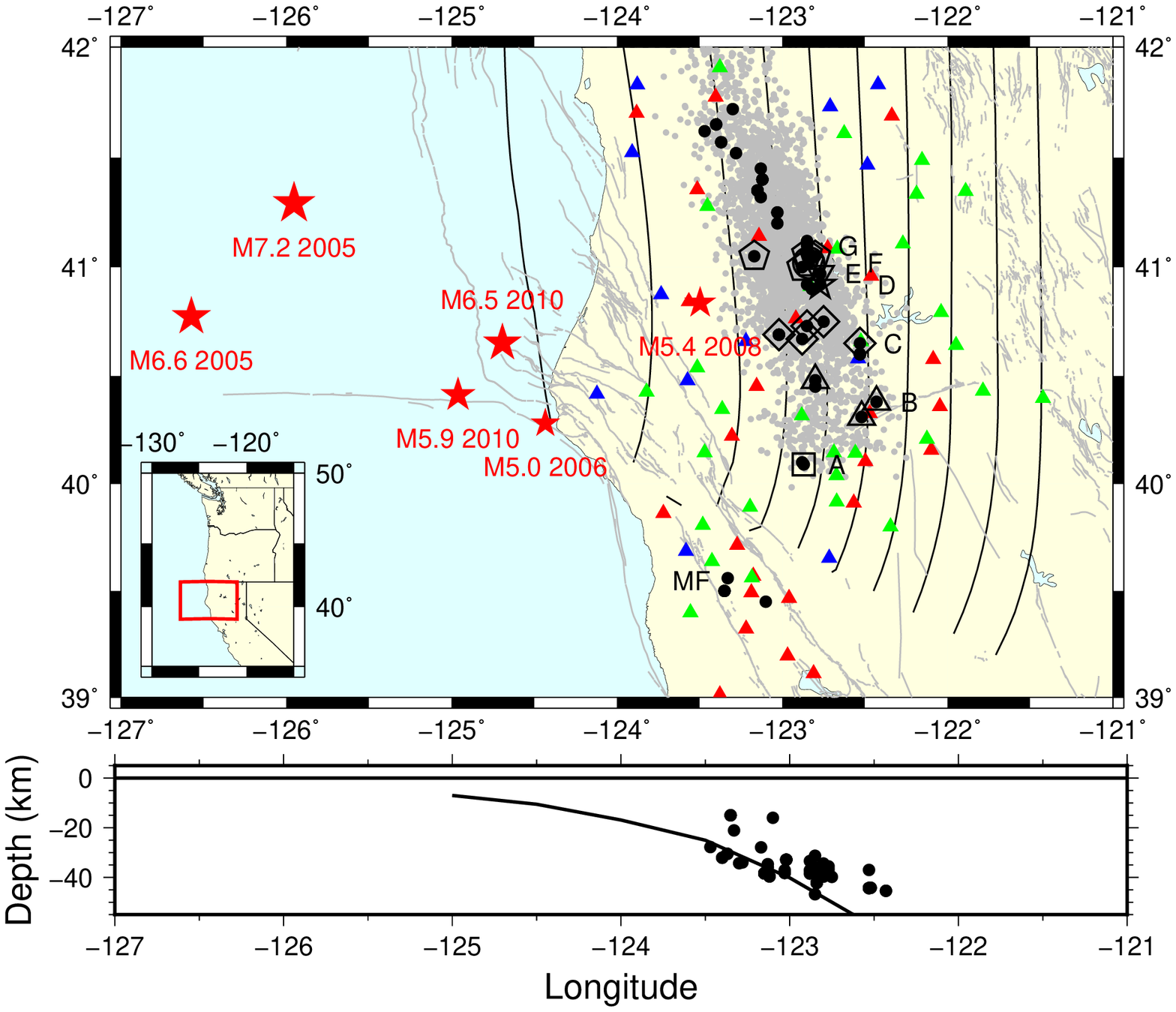}
\caption{Map showing the location of the LFE families (black dots) and the seismic stations used in this study. Red triangles are the stations from the FAME experiment, green triangles are one-component permanent stations, blue triangles are three components permanent stations. Red stars are moderate ($M > 5$) nearby earthquakes. The plate boundary from ~\citeA{MCC_2006} is contoured at 10-km intervals starting at 10 km depth. The cross section shows the depth of the LFE families ~\cite{PLO_2015} and the plate boundary model from ~\citeA{MCC_2006} at latitude 41 N. MF indicates the Maacama Fault. We highlighted with different shapes the LFE families or groups of LFE families that are discussed later in the text. The black square is for family A, the black triangles are for families B1 to B3, the black diamonds are for families C1 to C5, the black star is for family D, the inverted triangle is for family E, the circle is for family F, and the pentagons are for families G1 to G5.}
\label{pngfiguresample}
\end{figure}

~\citeA{PLO_2015} have kindly shared their LFE catalog with us. For each of 66 LFE families they provided hypocentral locations, lists of stations and channels used to detect LFEs, and the timing of all LFE detections. ~\citeA{PLO_2015} have later reduced the number of LFE families to 37, by grouping together families with many common detections, but we chose to use the initial detections to extend the catalog. Using this dataset, we have created LFE templates for each LFE family and each seismic station and channel. For a given LFE family, a given station and a given channel, we downloaded an 80-second-long seismic waveform starting 10 seconds before the LFE detection time, we detrended the data, tapered the first and last 5 seconds of the data with a Hann window, removed the instrument response, bandpassed filter between 1.5 and 9 Hz, resampled the data to 20 Hz, and cut the first and last 10 seconds of data to obtain a one-minute-long template. We then linearly stacked all the waveforms after normalizing each waveform with the root mean square (RMS) to obtain a waveform template for each station and each channel.

\section{Method}

Our ultimate goal is to obtain an eight-year-long LFE catalog using data from permanent seismic networks. The first step is to use the templates from ~\citeA{PLO_2015} and the data from the FAME experiment to get an LFE catalog for 2007-2009. The second step is to use the 2007-2009 FAME catalog to make templates for the seismic stations from the permanent networks. The third step is to use the new templates and the data from the permanent networks to get an LFE catalog for 2004-2011. We used a matched-filter algorithm to detect LFEs. For a given LFE family, we download one hour of seismic data. Then for each station and each channel, we cross-correlate the one-hour long signal with the one-minute-long template for the given station and channel. We scale the correlation coefficient by the standard deviation so the value is 1 if the correlation is perfect. As the signal-to-noise ratio of the seismic data is low, we may not see obvious peaks in the cross-correlation signal. However, if we stack the cross-correlation signals for all the channels and all the stations, we can see peaks appearing. Whenever the value of the average cross-correlation is higher than a threshold (we chose a threshold equal to eight times the median absolute deviation (MAD) of the stacked cross-correlation as was done by ~\citeA{SHE_2007_nature}), we inferred that there is an LFE. As two peaks separated by a short period of time may actually correspond to the same LFE, we kept only LFEs that are separated by at least one second and, when two LFEs are separated by less than one second, we keep only the one with the higher value of the stacked cross-correlation. \\

We first looked for LFEs during the months of March and April 2008, which correspond to the period covered by the catalog from ~\citeA{PLO_2015}, and compared our detections with the initial detections from the original catalog of 66 families. For 19 families, we recovered all the LFE detections initially present in the ~\citeA{PLO_2015} catalog. For 61 families, we recovered more than 90 \% of the initial LFE detections. We also added 74 \% more LFEs to the initial number of LFEs in the catalog (when considering all 66 families), but most of added LFEs have a low cross-correlation value and may be false detections. Figure 2 shows the cross-correlation values for the LFEs in the original catalog, and for the LFEs that we added to the catalog for family E (inverted triangle in Figure 1). \\

\begin{figure}
\noindent\includegraphics[width=\textwidth, trim={0cm 0cm 25.5cm 0cm},clip]{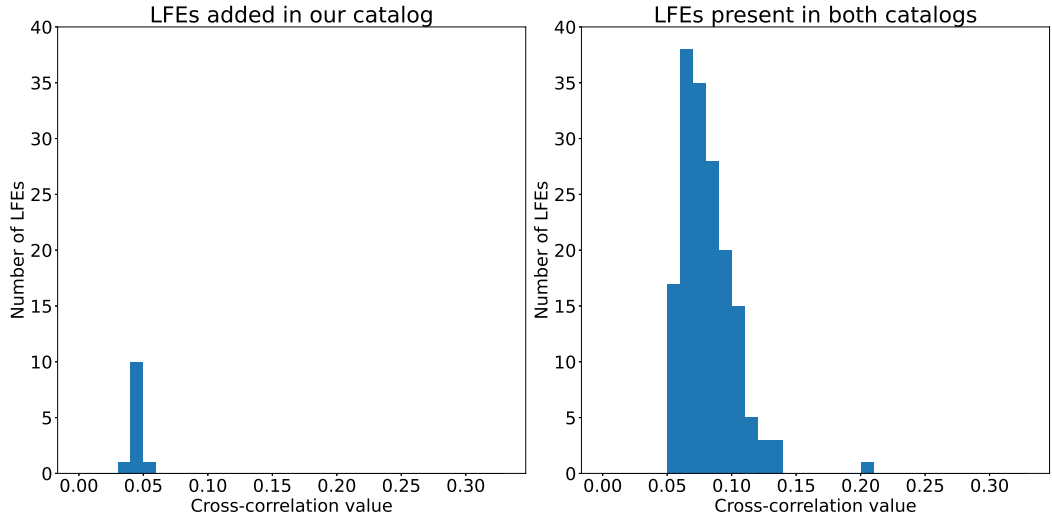}
\caption{Cross-correlation values for LFEs added in the catalog for family E (left) and cross-correlation values for LFEs in the original catalog from ~\citeA{PLO_2015} (right). There were no missing LFEs for this family.}
\label{pngfiguresample}
\end{figure}

We then looked for LFEs during the period from July 2007 to June 2009, which correspond to the period when the FAME experiment was operating. Using a threshold equal to eight times the MAD of the stacked cross-correlation may produce false detections, therefore we filtered the LFE detection times before visualizing the two-year-long catalog. Because the number of seismic stations recording may change with time as the stations were progressively installed during Summer and Fall 2007, and then progressively removed during May and June 2009, we kept only LFE detections for which the product of the stacked cross-correlation value by the number of channels recording at that time is higher than a threshold. We will explain later in the text how the threshold is chosen. The threshold is different for each LFE family, but is constant over time for a given LFE family. The resulting LFE catalog for the period 2007-2009 is shown in Figure 3. For comparison, we also plotted the tremor detection times from ~\citeA{BOY_2015}. Filtering the catalog has removed about 84 \% of the LFEs, compared to the initial catalog where LFEs were detected using the 8 * MAD threshold for the cross-correlation. As a reference, we plotted in the Supplementary Material (Figure S1) the catalog without filtering the LFEs and the number of channels recording as a function of time (Figure S3). \\

\begin{figure}
\noindent\includegraphics[width=\textwidth, trim={0cm 0cm 0cm 0cm},clip]{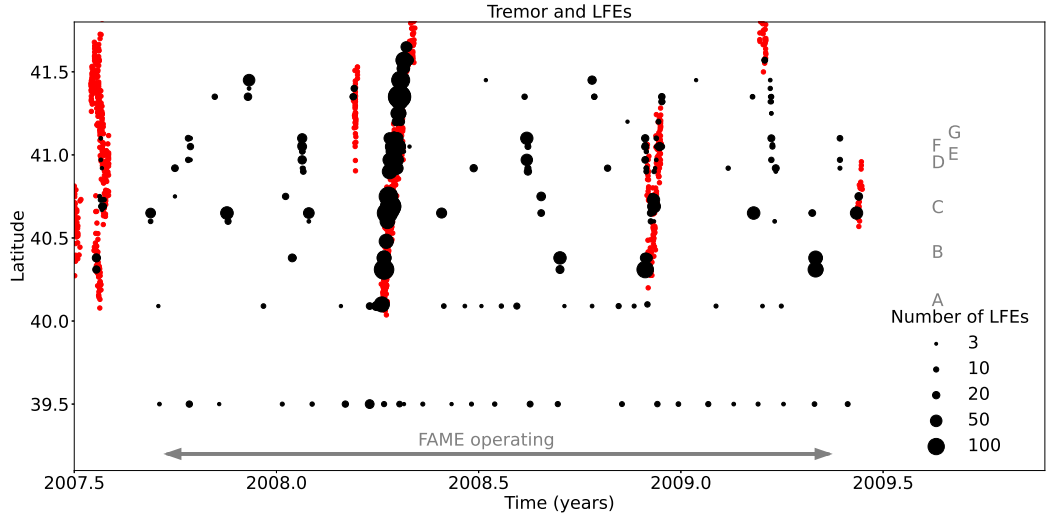}
\caption{Tremor and LFEs detected using the data from the FAME experiment as a function of time and latitude. Red dots represent tremor detections from the catalog of ~\citeA{BOY_2015}. Black dots represent days when LFEs are detected for a given LFE family. The size of each black dot is proportional to the number of LFEs detected during this day. The double-headed grey arrow represents the time period when the FAME experiment was operating at full capacity. LFE families south of 40° N latitude are on the San Andreas fault system. For clarity, we do not show the LFEs for the days when there are less than three LFEs.}
\label{pngfiguresample}
\end{figure}

We note that there is a good spatial and temporal agreement between tremor and LFEs, with LFEs detected during the main tremor episodes. Additional small LFE episodes are also detected between bigger tremor episodes. The LFE families located south on the strike-slip fault from the San Andreas Fault system have much shorter recurrence intervals than the LFE families located on the subduction zone. Additionally, one LFE family located on the southern end of the subduction zone (family A at 40.09N and 34.9km depth) also has shorter recurrence intervals than families located farther north, and behaves more similarly to the strike-slip fault families. \\

We then used the LFE detections from the 2007-2009 catalog to make new templates for the permanent stations of the three seismic networks: Berkeley Digital Seismic Network (BK), Northern California Seismic Network (NC), and Plate Boundary Observatory Strain and Seismic Data (PB). For a given LFE family, we took the 150 LFE detection times with the best cross-correlation value, we downloaded one minute of seismic data around each detection, and linearly stacked the waveforms to obtain the templates. We looked for templates for both one-component stations and three-component stations. \\

For most families, we find that we can obtain good templates with high signal-to-noise ratio for several stations. Only nine families have four or less seismic stations with good templates. Examples of templates are given in Figure 4 for station WDC and family A (square in Figure 1). To obtain the above catalog, we used one-minute-long templates, which included noise before and after the seismic wave arrivals. To increase the cross-correlation values between the templates and the data, we thus reduced the length of the templates to 25 to 40 seconds, depending on the maximum distance from the source to the stations. The 2007-2009 FAME catalog has thus been established with cross correlations over one-minute-long templates while the 2004-2011 networks catalog has been established with cross correlations over shorter templates with average duration equal to 30 seconds. We then used these new, shorter templates for the permanent stations to make an LFE catalog for the period 2007-2009. \\

\begin{figure}
\noindent\includegraphics[width=\textwidth, trim={0cm 0cm 0cm 0cm},clip]{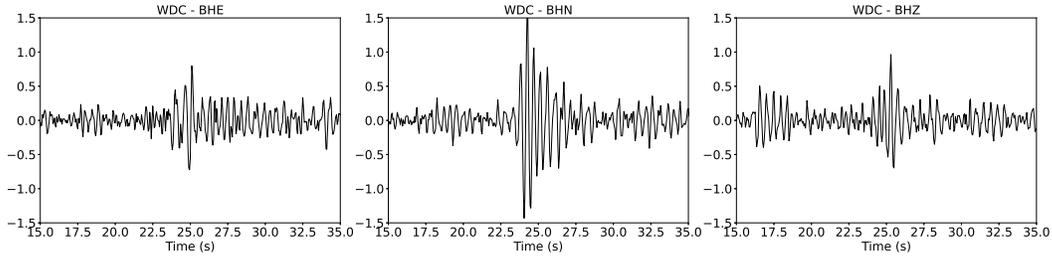}
\caption{Low-frequency earthquake template for station WDC and family A for the three channels BHE, BHN and BHZ (left to right). We can clearly see a P-wave arrival and an S-wave arrival about 7 seconds later. The P-wave has a higher amplitude on the vertical component, and the S-wave has a higher amplitude on the horizontal components.}
\label{pngfiguresample}
\end{figure}


Once we have obtained a catalog for the period 2007-2009 using the data from the permanent seismic networks, we compare the LFE detections between the two catalogs: the FAME catalog (obtained with data from the FAME experiment) and the networks catalog (obtained from data from the permanent networks). As we may have many false detections, we try to eliminate some of them by assuming that LFEs present in both catalogs are always true detections. Then we define two thresholds for each LFE family: the first threshold is chosen such that half the LFE detections above the threshold in the FAME catalog are also in the network catalog, the second threshold is chosen such that half the LFE detections above the threshold (and occurring during 2007-2009) in the network catalog are also in the FAME catalog. If we filter the catalogs and keep only detections above the thresholds, we are now confident that at least half of the LFE detections are true detections. Thus there are two thresholds per LFE family (given in Table S3 of the Supplementary Material). As the number of stations recording may vary over time (especially for the permanent networks catalog), we keep only the LFE detections such that the associated cross correlation multiplied by the number of channels recording at that time is higher than the threshold. Thus, if there are few stations recordings at some time, the cross correlation must be higher for the LFE to be considered as a true detection. We then compared the normalized number of LFEs obtained with the two catalogs. Examples are given in Figure 5 for families A (square in Figure 1) and E (inverted triangle in Figure 1). In the following, we speak of an LFE cluster when many LFEs are detected for a given family in a short period of time, we speak of LFE episode when many LFEs are detected for several families and LFE activity is propagating from one family to another along the strike of the plate boundary. Although the number of LFEs during each cluster of LFEs may not be the same, we clearly see for family E that the timing of the clusters are the same for both catalogs. Family A has much shorter recurrence times but most clusters of LFEs are present in both catalogs as well: forty clusters are present in both catalogs, three clusters are present in the FAME catalog but not in the networks catalog and one cluster is present in the networks catalog but not in the FAME catalog. \\

\begin{figure}
\noindent\includegraphics[width=\textwidth, trim={0cm 0cm 0cm 0cm},clip]{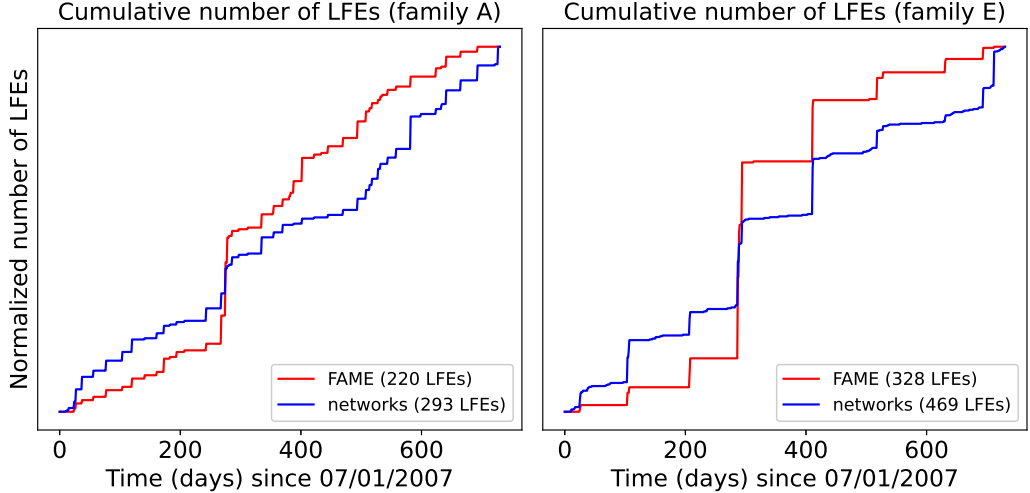}
\caption{Normalized cumulative number of LFEs over the period July 2007-July 2009 for the FAME catalog (red) and the network catalog (blue) for family A and family E. The cumulative number of LFEs has been divided by the total number of LFEs so that both curves start at 0 and end at 1, in order to get a better match between the LFE clusters in both catalogs.}
\label{pngfiguresample}
\end{figure}

For most of the families located on the subduction zone, we obtained a good agreement between both catalogs. We eliminate six families for which too few permanent stations have good templates, and it was not possible to detect LFEs with the available templates. However, we were able to detect most LFE clusters even without the presence of the temporary stations from the FAME network. We are thus confident that we can detect LFEs before 2007 and after 2009. In the following, we focus on the period 2004-2011. In November 2011, several one-component stations stopped recording and the number of available stations started decreasing, which is why we did not look for LFEs after that date. The resulting LFE catalog for the period 2004-2011 is shown in Figure 6. For comparison, we also plotted the tremor detection times from ~\citeA{BOY_2015}. Figure 7 shows the details of six big ETS events in 2004, 2005, 2006, 2007 and 2008. The catalog shown in Figure 6 has been filtered using the threshold computed above. Filtering the catalog has removed about 63 \% of the LFEs, compared to the initial catalog where LFEs were detected using the 8 * MAD threshold for the cross-correlation. As a reference, we plotted in the Supplementary Material (Figure S2) the catalog without filtering the LFEs and the number of channels recording as a function of time (Figure S4). \\

\begin{figure}
\noindent\includegraphics[width=\textwidth, trim={0cm 0cm 0cm 0cm},clip]{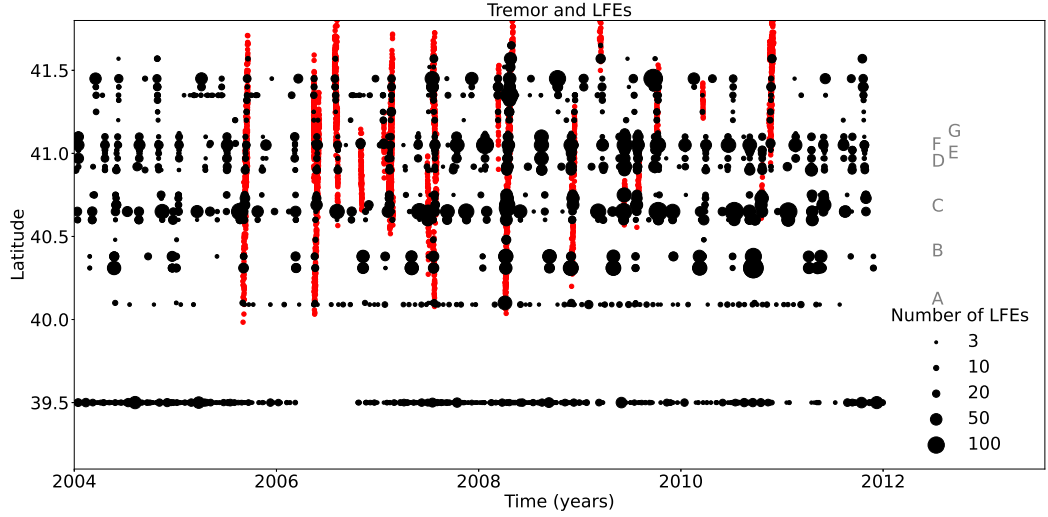}
\caption{Tremor and LFEs detected using the data from the permanent seismic networks as a function of time and latitude. Red dots represent tremor detections from the catalog of ~\citeA{BOY_2015}. Black dots represent days when LFEs are detected for a given LFE family. The size of each black dot is proportional to the number of LFEs detected during this day. For clarity, we do not show the LFEs for the days when there are less than three LFEs.}
\label{pngfiguresample}
\end{figure}

\begin{figure}
\noindent\includegraphics[width=\textwidth, trim={0cm 0cm 0cm 0cm},clip]{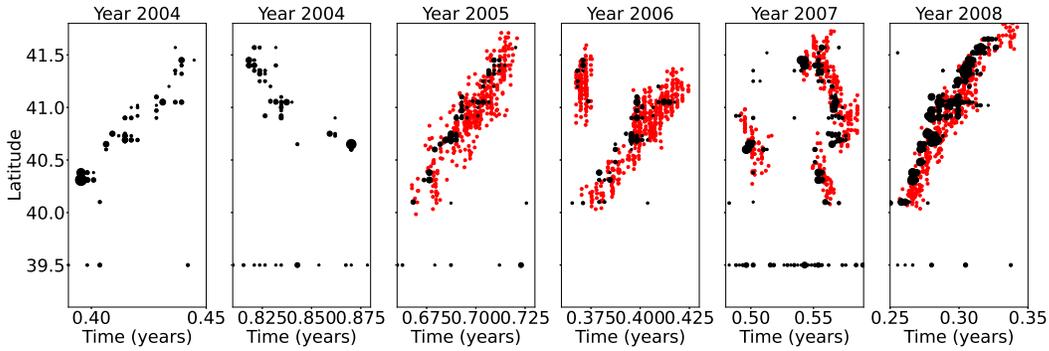}
\caption{LFE and tremor detections as a function of time and latitude for six ETS events. Red dots represent tremor detections from the catalog of ~\citeA{BOY_2015}. Black dots represent days when LFEs are detected for a given LFE family. The size of each black dot is proportional to the number of LFEs detected during this day. For clarity, we do not show the LFEs for the days when there are less than three LFEs. The tremor catalog of ~\citeA{BOY_2015} starts in 2005. So there are no tremor detections for the two ETS events in 2004.}
\label{pngfiguresample}
\end{figure}

\section{Discussion}

We were able to detect LFE episodes propagating from North to South or from South to North whenever there is a tremor episode in the ~\citeA{BOY_2015} catalog. We were also able to detect several LFE episodes in 2004-2005 before the beginning of the ~\citeA{BOY_2015} catalog. In particular, the LFE episode in January 2005 propagating South to North from latitude 40.1 to latitude 41.2 is also present in the ETS catalog from ~\citeA{BRU_2007}. We also see one LFE episode in May-June 2004, propagating South to North from latitude 40.1 to 41.6, and one LFE episode in November 2004, propagating North to South from latitude 41.6 to 40.4. The first of these episodes is also in the ETS catalog from ~\citeA{BRU_2007}. The main characteristics of the six biggest episodes shown in Figure 7 are given in Table 1. \\

\begin{table}
\caption{Main LFEs episodes between 2004 and 2011}
\centering
\begin{tabular}{c c c c c}
\hline
Begin time & End time & Direction & Southern limit & Northern limit \\
\hline
End May 2004 & Mid June 2004 & South to North & 40.1 & 41.6 \\
Early November 2004 & End November 2004 & North to South & 40.4 & 41.6 \\
Early September 2005 & End September 2005 & South to North & 40.1 & 41.6 \\
Mid April 2006 & End April 2006 & South to North & 40.1 & 41.5 \\
Early July 2007 & End July 2007 & North to South & 40.1 & 41.6 \\
Early April 2008 & End April 2008 & South to North & 40.1 & 41.7 \\
\hline
\end{tabular}
\end{table}

It is known that tectonic tremor can be triggered by surface waves from distant and regional earthquakes, as has been observed in Cascadia ~\cite{RUB_2009}, the San Andreas ~\cite{PEN_2009, GUI_2010}, and Nankai ~\cite{MIY_2008,HAN_2014}. LFE activity on the San Andreas fault also increased during several months after the 2004 Parkfield earthquake ~\cite{SHE_2017}. We looked for regional earthquakes with magnitude higher than 5 during the period covered by the catalog to verify whether this phenomenon is also observed in southern Cascadia. The characteristics of the earthquakes we looked at are given in Table 2. We also looked at five large, distant earthquakes (Table 3). \\

\begin{table}
\caption{Nearby regional earthquakes between 2004 and 2011}
\centering
\begin{tabular}{c c c c c}
\hline
Magnitude & Time & Latitude & Longitude & Depth (km) \\
\hline
7.2 & 2005-06-15 02:50:54 & 41.292N & 125.953W & 16.0 \\
6.6 & 2005-06-17 06:21:42 & 40.773N & 126.574W & 12.0 \\
6.5 & 2010-01-10 00:27:39 & 40.652N & 124.693W & 28.7 \\
5.9 & 2010-02-04 20:20:21 & 40.412N & 124.961W & 23.0 \\
5.4 & 2008-04-30 03:03:06 & 40.836N & 123.497W & 27.8 \\
5.0 & 2006-07-19 11:41:43 & 40.281N & 124.433W & 20.1 \\
\hline
\end{tabular}
\end{table}

\begin{table}
\caption{Teleseismic earthquakes between 2004 and 2011}
\centering
\begin{tabular}{c c c c c}
\hline
Magnitude & Time & Latitude & Longitude & Depth (km) \\
\hline
9.1 & 2011-03-11 05:46:24 & 38.297N & 142.373E & 29.0 \\
9.1 & 2004-12-26 00:58:53 & 3.295N & 95.982E & 30.0 \\
8.8 & 2010-02-27 06:34:11 & 36.122S & 72.898W & 22.9 \\ 
8.6 & 2005-03-28 16:09:36 & 2.085N & 97.108E & 30.0 \\ 
7.9 & 2008-05-12 06:28:01 & 31.002N & 103.322E & 19.0 \\ 
\hline
\end{tabular}
\end{table}

Although we observe an increase in LFE activity for some families several days after the earthquakes (except the 2008 one where no activity is observed after the earthquake), it does not seem that this activity is linked to the earthquake. Indeed, the LFE clusters occur a few days after the earthquake and not immediately after as is the case for the Parkfield earthquake, and they affect only a few families. The reason may be due to the distance between the LFE families and the epicenters of the regional earthquakes. Indeed, for the Parkfield earthquake, an increase in LFE activity was observed for LFE families up to 45km away from the epicenter, but not farther away. Moreover, the Parkfield earthquake was more shallow (8.1km deep) than the regional earthquakes used in this study. For the 2003 M6.5 San Simeon earthquake, no increase in LFE activity in the San Andreas LFE families was observed in the hours following the earthquake. This was also a shallow earthquake (8.4km deep) but the epicenter was farther away from the LFE families (60 to 100 km). For southern Cascadia, the closest epicenter (2008) is located 50 km away from the closest LFE families. Based on what was observed for the San Andreas LFE families, this distance may be too large for a regional earthquake to trigger an increase in LFE activity. We do not see any change in LFE activity after the large teleseismic earthquakes. There was not any change in LFE activity for the San Andreas LFE families either. \\

In northern Cascadia, there is an increase in activity downdip of the plate boundary compared to up dip, both for tremor ~\cite{WEC_2011} and LFEs ~\cite{SWE_2019}. We want to verify whether we can see an increase of LFE activity for the downdip LFE families compared to the updip LFE families. Most families in the northern and central part of the study area are well aligned along the strike direction, however families in the southern part of the subduction zone are more distributed in the dip direction. For example, we consider 3 groups of LFE families labeled as B, C and G (triangles, diamonds and pentagons in Figure 1). Each group has 3-5 families at about the same along-strike position, but varying in the dip direction. For each of the three groups there is a very clear pattern of many more LFE clusters for the down-dip (eastern) families than for the updip (western) families (Figure 8), as was also the case in northern Cascadia ~\cite{SWE_2019}. It is more difficult to compare cluster size and swarm duration as the number of stations and the quality of the templates are different for each family, and may have a strong influence on the number of LFEs detected. To summarize the findings, we plot in Figure 9 the average time between LFE clusters as a function of the distance in the eastern direction from the western limit of the tremor. We define an LFE cluster as at least five LFEs recorded during a given day for a given LFE family. If two clusters are separated by less than five days, we assume that this is a single cluster. We clearly see that the average time between LFE clusters decreases with the down dip distance. The southernmost family A is also plotted in Figure 9, but does not fit the general trend.\\

\begin{figure}
\noindent\includegraphics[width=\textwidth, trim={0cm 0cm 0cm 0cm},clip]{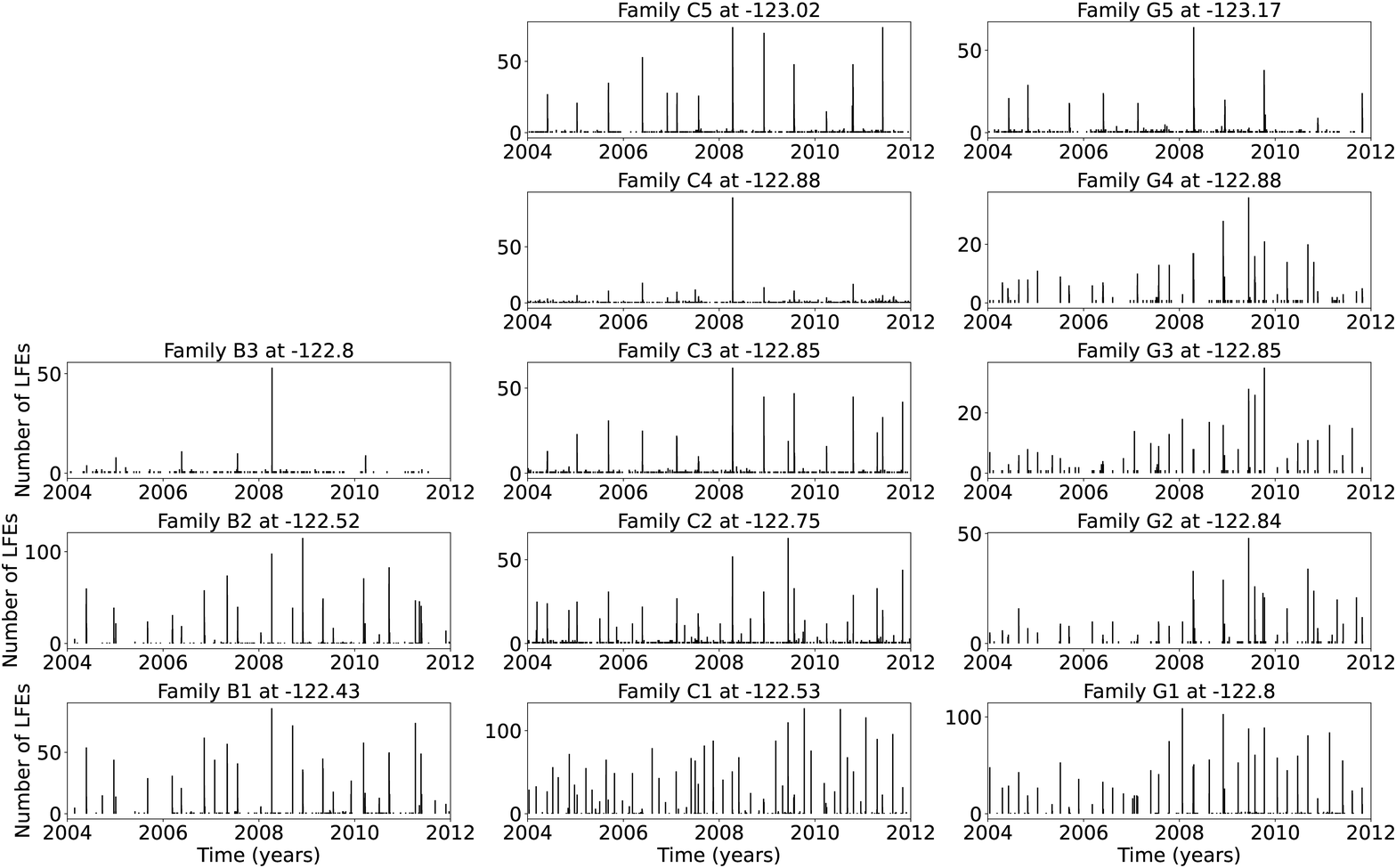}
\caption{Daily number of LFEs over the period 2004-2011 for the three LFE families in group B (left in this figure, triangles in Figure 1), for the five LFE families in group C (middle in this figure, diamonds in Figure 1) and the five LFE families in group G (right in this figure, pentagons in Figure 1). Down-dip families (lower rows) typically have far more LFE swarms that up-dip families (upper rows).}
\label{pngfiguresample}
\end{figure}

\begin{figure}
\noindent\includegraphics[width=\textwidth, trim={0cm 0cm 0cm 0cm},clip]{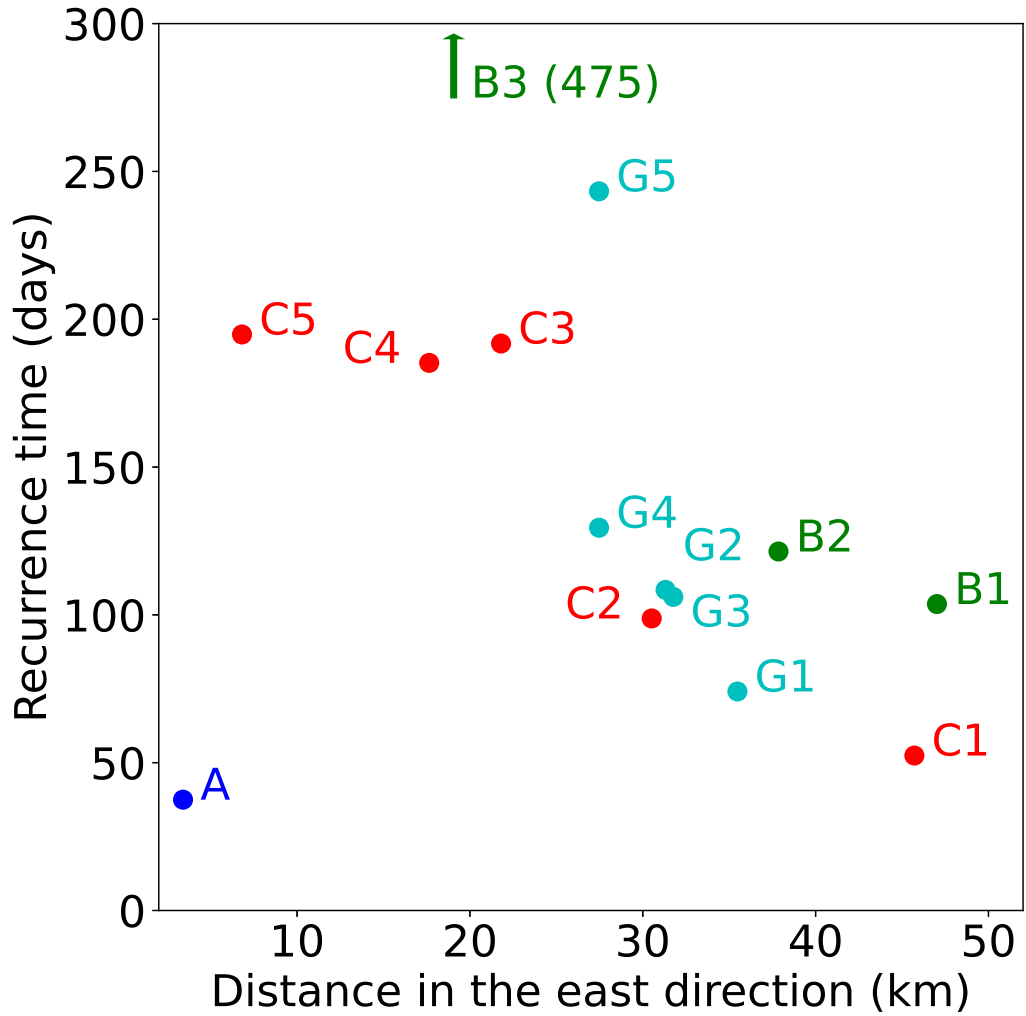}
\caption{Average time between LFE clusters as a function of distance in the eastern direction from the westward limit of tremor for the three LFE families in group B, the five LFE families in group C, and the five LFE families in group G. Family A has much shorter time intervals between LFE clusters.}
\label{pngfiguresample}
\end{figure}

The occurrence and amplitude of tectonic tremor can be modulated by small stress changes (about 1 kPa) associated with ocean or solid earth tides ~\cite{SHE_2007_G3,HOU_2011,HOU_2015}. Sensitivity of LFE families activity to tidal stress changes has also been observed in northern Cascadia ~\cite{ROY_2015}, and on the San Andreas fault ~\cite{THO_2012}. We used calculated stresses from elastic deformations due to both water loads and body tides in the solid Earth along the Cascadia subduction zone. The stress computation is done following the method described in ~\citeA{HOU_2015}. The resulting full stress tensors are projected to normal and shear (in the plate convergence direction) stresses on the interface, as well as mean stress (one-third the tensor trace). The change in Coulomb stress (positive promotes shear failure) $\Delta CFF$ is then computed:

\begin{linenomath*}
\begin{equation}
\Delta CFF = \Delta \tau + \mu \left( \Delta \sigma - B \Delta \sigma_m \right)
\end{equation}
\end{linenomath*}

where $\Delta \tau$ is the change in shear stress on the fault plane in the slip direction, $\Delta \sigma$ is the change in fault normal stress (positive is tensile), $\mu$ is the coefficient of intrinsic friction, $\Delta \sigma_m$ is the change in mean stress and $B$ is the Skempton's coefficient ~\cite{HOU_2015}. We take the values $\mu = 0.1$ and $B = 0.5$ for the friction and the Skempton's coefficient, as was done by ~\citeA{HOU_2015}. We then divided the values of tidal Coulomb stress change into 20 stress bins and computed for each LFE family the number of LFEs occurring at a time when the value of the stress was in a given bin. We also computed what would be the expected number of LFEs occurring if the tidal stress changes have no influence on LFE activity. As noted by ~\citeA{HOU_2015}, tremor sensitivity to stress perturbations changes during an ETS event, with an increase in sensitivity to tidal stress as slip accumulates over several days. We thus choose to only consider the later part of the 2008 ETS event, using only LFEs that occurred between 1.5 and 11.5 days after the arrival of the tremor front. As there are only a few LFEs in many of the families, the influence of tidal stress on LFE activity is not always clear. However, Figure 10 shows the number of LFEs occurring during each stress bin (blue bars) and the expected number of LFEs (black line) for six families with a relatively high number of LFEs. For most of the families, a positive Coulomb stress change (promoting shear failure) is associated with an increase in LFE activity, as is the case in northern Cascadia ~\cite{HOU_2015}. \\

However, this is not the case for the southernmost LFE family A, for which a negative Coulomb stress change is associated with an increase in LFE activity. To better highlight the behavior of family A, we also consider the whole LFE catalog and plot the observed and expected number of LFEs as a function of tidal stress. We clearly see that these LFEs tend to occur during times when predicted tidal Coulomb stress is discouraging slip on the plate boundary. We tried to relocate this LFE family using new templates obtained from LFEs present in both the 2007-2009 FAME catalog and the 2004-2011 networks catalog. We locate the family using S minus P times for each station, or using P-P times between pairs of stations, or using a combination of both. The family may be located 30 to 50 kilometers east from the initial location given by ~\citeA{PLO_2015}. The depth is 20-25 kilometers, that is shallower than the location given by ~\citeA{PLO_2015} and shallower than the plate boundary. However, this does not change the pattern observed for the correlation between tidal stress changes and LFE activity. We hypothesize that LFE family A may not be located on the subduction zone and may be on a nearby crustal fault. If this was the case, the fault plane assumed for the tidal stress calculation would not be appropriate, which could explain the different pattern for this LFE family. The LFE families located on the San Andreas fault by ~\citeA{SHE_2017} have a depth of 15 to 30 kilometers, so the depth of the LFE family A is compatible with a location on a crustal fault.

\begin{figure}
\noindent\includegraphics[width=\textwidth, trim={0cm 0cm 0cm 0cm},clip]{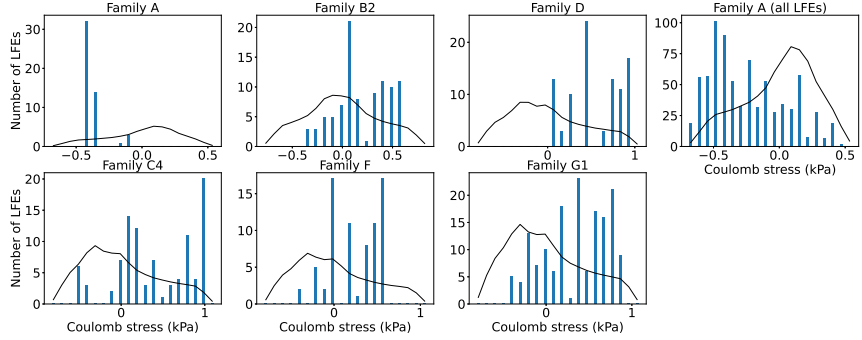}
\caption{Number of LFEs for families A, B2, D, C4, F and G1 observed (blue bars) and expected (black line) for each stress bin of tidal Coulomb stress change.}
\label{pngfiguresample}
\end{figure}

\section{Conclusion}

We used the waveform templates obtained by ~\citeA{PLO_2015} using data recorded by the FAME network in Northern California during an episode of high tremor activity in April 2008, and we extended their catalog to the whole two years (2007-2009) during which the seismic stations were installed. We then used the two-year-long catalog to create templates for stations from the permanent seismic networks, and used the seismic data recorded by these stations to extend the catalog before and after the FAME experiment to an eight-year duration (2004-2011). We observe abundant LFEs every time that there is tectonic tremor on the tremor catalog from Boyarko et al. (2015), and several additional smaller LFE episodes as well as two new longer LFE episodes in 2004 that predate the ~\citeA{BOY_2015} catalog. The recurrence intervals of LFE clusters are systematically longer for updip (west) families than for families that are downdip (east), as has been observed in Northern Cascadia ~\cite{SWE_2019}. LFEs occur far more frequently during encouraging tidal Coulomb stresses than during times of discouraging stress during the April 2008 ETS. The effect is strongest starting 1.5 days after the rupture front has passed a given family. The behavior has been seen in northern Cascadia as well ~\cite{HOU_2015}. The southernmost LFE family, near the southern edge of the slab is unusual in that it is updip, yet has the shortest recurrence interval of all the families, and it occurs much more often during discouraging tidal stresses than during encouraging stress. Perhaps this family is mislocated and is on a crustal strike-slip fault, or on the edge of the slab rather than on the subduction interface.

\acknowledgments
The authors would like to thank A. Plourde for sharing his low-frequency earthquake catalog, D. Boyarko for sharing his tremor catalog, and H. Houston for the tidal stress calculation in the Cascadia subduction zone. The authors would like to thank Alexandre Plourde, an anonymous reviewer and the associate editor whose comments helped improve the manuscript. This project was funded by NSF grant EAR-1358512. A.D. would like to thank the Integral Environmental Big Data Research Fund from the College of the Environment of University of Washington, which funded cloud computing resources to carry out the data analyses. The seismic recordings from the FAME experiment used for this analysis can be downloaded from the IRIS website \cite{https://doi.org/10.7914/sn/xq_2007}. Waveform data from the permanent networks were accessed through the Northern California Earthquake Data Center (NCEDC), doi:10.7932/NCEDC. The first figure was done using GMT ~\cite{WES_1991}. All the downloading and preprocessing operations on the seismic waveforms are done with the Python package obspy. The Python scripts used to analyze the data and make the figures can be found on the first author's Github account, accessible through Zenodo ~\cite{ariane_ducellier_2022_6374239}.


%
%

\bibliography{bibliography}

\begin{thebibliography}{}

\bibitem [\protect \citeauthoryear {%
Audet%
\ \BBA {} Kim%
}{%
Audet%
\ \BBA {} Kim%
}{%
{\protect \APACyear {2016}}%
}]{%
AUD_2016}
\APACinsertmetastar {%
AUD_2016}%
\begin{APACrefauthors}%
Audet, P.%
\BCBT {}\ \BBA {} Kim, Y\BPBI H.%
\end{APACrefauthors}%
\unskip\
\newblock
\APACrefYearMonthDay{2016}{}{}.
\newblock
{\BBOQ}\APACrefatitle {Teleseismic constraints on the geological environment of
  deep episodic slow earthquakes in subduction zone forearcs: A review}
  {Teleseismic constraints on the geological environment of deep episodic slow
  earthquakes in subduction zone forearcs: A review}.{\BBCQ}
\newblock
\APACjournalVolNumPages{Tectonophysics}{670}{}{1-15}.
\PrintBackRefs{\CurrentBib}

\bibitem [\protect \citeauthoryear {%
Baratin%
, Chamberlain%
, Townend%
\BCBL {}\ \BBA {} Savage%
}{%
Baratin%
\ \protect \BOthers {.}}{%
{\protect \APACyear {2018}}%
}]{%
BAR_2018}
\APACinsertmetastar {%
BAR_2018}%
\begin{APACrefauthors}%
Baratin, L\BPBI M.%
, Chamberlain, C\BPBI J.%
, Townend, J.%
\BCBL {}\ \BBA {} Savage, M\BPBI K.%
\end{APACrefauthors}%
\unskip\
\newblock
\APACrefYearMonthDay{2018}{}{}.
\newblock
{\BBOQ}\APACrefatitle {Focal mechanisms and inter-event times of low-frequency
  earthquakes reveal quasi-continuous deformation and triggered slow slip on
  the deep {Alpine} {Fault}} {Focal mechanisms and inter-event times of
  low-frequency earthquakes reveal quasi-continuous deformation and triggered
  slow slip on the deep {Alpine} {Fault}}.{\BBCQ}
\newblock
\APACjournalVolNumPages{Earth and Planetary Science Letters}{484}{}{111-123}.
\PrintBackRefs{\CurrentBib}

\bibitem [\protect \citeauthoryear {%
Beroza%
\ \BBA {} Ide%
}{%
Beroza%
\ \BBA {} Ide%
}{%
{\protect \APACyear {2011}}%
}]{%
BER_2011}
\APACinsertmetastar {%
BER_2011}%
\begin{APACrefauthors}%
Beroza, G\BPBI C.%
\BCBT {}\ \BBA {} Ide, S.%
\end{APACrefauthors}%
\unskip\
\newblock
\APACrefYearMonthDay{2011}{}{}.
\newblock
{\BBOQ}\APACrefatitle {Slow earthquakes and nonvolcanic tremor} {Slow
  earthquakes and nonvolcanic tremor}.{\BBCQ}
\newblock
\APACjournalVolNumPages{Annu. Rev. Earth Planet. Sci.}{39}{}{271-296}.
\PrintBackRefs{\CurrentBib}

\bibitem [\protect \citeauthoryear {%
Bostock%
, Royer%
, Hearn%
\BCBL {}\ \BBA {} Peacock%
}{%
Bostock%
\ \protect \BOthers {.}}{%
{\protect \APACyear {2012}}%
}]{%
BOS_2012}
\APACinsertmetastar {%
BOS_2012}%
\begin{APACrefauthors}%
Bostock, M\BPBI G.%
, Royer, A\BPBI A.%
, Hearn, E\BPBI H.%
\BCBL {}\ \BBA {} Peacock, S\BPBI M.%
\end{APACrefauthors}%
\unskip\
\newblock
\APACrefYearMonthDay{2012}{}{}.
\newblock
{\BBOQ}\APACrefatitle {Low frequency earthquakes below southern {Vancouver}
  {Island}} {Low frequency earthquakes below southern {Vancouver}
  {Island}}.{\BBCQ}
\newblock
\APACjournalVolNumPages{Geochemistry Geophysics Geosystems}{13}{}{Q11007}.
\PrintBackRefs{\CurrentBib}

\bibitem [\protect \citeauthoryear {%
Bostock%
, Thomas%
, Savard%
, Chuang%
\BCBL {}\ \BBA {} Rubin%
}{%
Bostock%
\ \protect \BOthers {.}}{%
{\protect \APACyear {2015}}%
}]{%
BOS_2015}
\APACinsertmetastar {%
BOS_2015}%
\begin{APACrefauthors}%
Bostock, M\BPBI G.%
, Thomas, A\BPBI M.%
, Savard, G.%
, Chuang, L.%
\BCBL {}\ \BBA {} Rubin, A\BPBI M.%
\end{APACrefauthors}%
\unskip\
\newblock
\APACrefYearMonthDay{2015}{}{}.
\newblock
{\BBOQ}\APACrefatitle {Magnitudes and moment-duration scaling of low-frequency
  earthquakes beneath southern {Vancouver} {Island}} {Magnitudes and
  moment-duration scaling of low-frequency earthquakes beneath southern
  {Vancouver} {Island}}.{\BBCQ}
\newblock
\APACjournalVolNumPages{Journal of Geophysical Research: Solid
  Earth}{120}{}{6329-6350}.
\PrintBackRefs{\CurrentBib}

\bibitem [\protect \citeauthoryear {%
Boyarko%
, Brudzinski%
, Porritt%
, Allen%
\BCBL {}\ \BBA {} Tr\'ehu%
}{%
Boyarko%
\ \protect \BOthers {.}}{%
{\protect \APACyear {2015}}%
}]{%
BOY_2015}
\APACinsertmetastar {%
BOY_2015}%
\begin{APACrefauthors}%
Boyarko, D\BPBI C.%
, Brudzinski, M\BPBI R.%
, Porritt, R\BPBI W.%
, Allen, R\BPBI M.%
\BCBL {}\ \BBA {} Tr\'ehu, A\BPBI M.%
\end{APACrefauthors}%
\unskip\
\newblock
\APACrefYearMonthDay{2015}{}{}.
\newblock
{\BBOQ}\APACrefatitle {Automated detection and location of tectonic tremor
  along the entire {Cascadia} margin from 2005 to 2011} {Automated detection
  and location of tectonic tremor along the entire {Cascadia} margin from 2005
  to 2011}.{\BBCQ}
\newblock
\APACjournalVolNumPages{Earth and Planetary Science Letters}{430}{}{160-170}.
\PrintBackRefs{\CurrentBib}

\bibitem [\protect \citeauthoryear {%
Brudzinski%
\ \BBA {} Allen%
}{%
Brudzinski%
\ \BBA {} Allen%
}{%
{\protect \APACyear {2007}}%
}]{%
BRU_2007}
\APACinsertmetastar {%
BRU_2007}%
\begin{APACrefauthors}%
Brudzinski, M\BPBI R.%
\BCBT {}\ \BBA {} Allen, R\BPBI M.%
\end{APACrefauthors}%
\unskip\
\newblock
\APACrefYearMonthDay{2007}{}{}.
\newblock
{\BBOQ}\APACrefatitle {Segmentation in episodic tremor and slip all along
  {Cascadia}} {Segmentation in episodic tremor and slip all along
  {Cascadia}}.{\BBCQ}
\newblock
\APACjournalVolNumPages{Geology}{35}{10}{907-910}.
\PrintBackRefs{\CurrentBib}

\bibitem [\protect \citeauthoryear {%
Chestler%
\ \BBA {} Creager%
}{%
Chestler%
\ \BBA {} Creager%
}{%
{\protect \APACyear {2017}}%
{\protect \APACexlab {{\protect \BCnt {1}}}}}]{%
CHE_2017_JGR}
\APACinsertmetastar {%
CHE_2017_JGR}%
\begin{APACrefauthors}%
Chestler, S\BPBI R.%
\BCBT {}\ \BBA {} Creager, K\BPBI C.%
\end{APACrefauthors}%
\unskip\
\newblock
\APACrefYearMonthDay{2017{\protect \BCnt {1}}}{}{}.
\newblock
{\BBOQ}\APACrefatitle {Evidence for a scale-limited low-frequecny earthquake
  source process} {Evidence for a scale-limited low-frequecny earthquake source
  process}.{\BBCQ}
\newblock
\APACjournalVolNumPages{Journal of Geophysical Research: Solid
  Earth}{122}{}{3099-3114}.
\PrintBackRefs{\CurrentBib}

\bibitem [\protect \citeauthoryear {%
Chestler%
\ \BBA {} Creager%
}{%
Chestler%
\ \BBA {} Creager%
}{%
{\protect \APACyear {2017}}%
{\protect \APACexlab {{\protect \BCnt {2}}}}}]{%
CHE_2017_G3}
\APACinsertmetastar {%
CHE_2017_G3}%
\begin{APACrefauthors}%
Chestler, S\BPBI R.%
\BCBT {}\ \BBA {} Creager, K\BPBI C.%
\end{APACrefauthors}%
\unskip\
\newblock
\APACrefYearMonthDay{2017{\protect \BCnt {2}}}{}{}.
\newblock
{\BBOQ}\APACrefatitle {A model for low-frequency earthquake slip} {A model for
  low-frequency earthquake slip}.{\BBCQ}
\newblock
\APACjournalVolNumPages{Geochemistry, Geophysics, Geosystems}{18}{}{4690-4708}.
\PrintBackRefs{\CurrentBib}

\bibitem [\protect \citeauthoryear {%
Ducellier%
}{%
Ducellier%
}{%
{\protect \APACyear {2022}}%
}]{%
ariane_ducellier_2022_6374239}
\APACinsertmetastar {%
ariane_ducellier_2022_6374239}%
\begin{APACrefauthors}%
Ducellier, A.%
\end{APACrefauthors}%
\unskip\
\newblock
\APACrefYearMonthDay{2022}{{\APACmonth{03}}}{}.
\newblock
\APACrefbtitle {ArianeDucellier/catalog: Second revision.}
  {Arianeducellier/catalog: Second revision.}
\newblock
\APACaddressPublisher{}{Zenodo}.
\newblock
\begin{APACrefURL} \url{https://doi.org/10.5281/zenodo.6374239}
  \end{APACrefURL}
\newblock
\begin{APACrefDOI} \doi{10.5281/zenodo.6374239} \end{APACrefDOI}
\PrintBackRefs{\CurrentBib}

\bibitem [\protect \citeauthoryear {%
Frank%
\ \protect \BOthers {.}}{%
Frank%
\ \protect \BOthers {.}}{%
{\protect \APACyear {2014}}%
}]{%
FRA_2014}
\APACinsertmetastar {%
FRA_2014}%
\begin{APACrefauthors}%
Frank, W\BPBI B.%
, Shapiro, N\BPBI M.%
, Husker, A\BPBI L.%
, Kostoglodov, V.%
, Romanenko, A.%
\BCBL {}\ \BBA {} Campillo, M.%
\end{APACrefauthors}%
\unskip\
\newblock
\APACrefYearMonthDay{2014}{}{}.
\newblock
{\BBOQ}\APACrefatitle {Using systematically characterized low-frequency
  earthquakes as a fault probe in {Guerrero}, {Mexico}} {Using systematically
  characterized low-frequency earthquakes as a fault probe in {Guerrero},
  {Mexico}}.{\BBCQ}
\newblock
\APACjournalVolNumPages{Journal of Geophysical Research: Solid
  Earth}{119}{}{7686-7700}.
\PrintBackRefs{\CurrentBib}

\bibitem [\protect \citeauthoryear {%
Guilhem%
, Peng%
\BCBL {}\ \BBA {} Nadeau%
}{%
Guilhem%
\ \protect \BOthers {.}}{%
{\protect \APACyear {2010}}%
}]{%
GUI_2010}
\APACinsertmetastar {%
GUI_2010}%
\begin{APACrefauthors}%
Guilhem, A.%
, Peng, Z.%
\BCBL {}\ \BBA {} Nadeau, R\BPBI M.%
\end{APACrefauthors}%
\unskip\
\newblock
\APACrefYearMonthDay{2010}{}{}.
\newblock
{\BBOQ}\APACrefatitle {High-frequency identification of non-volcanic tremor
  triggered by regional earthquakes} {High-frequency identification of
  non-volcanic tremor triggered by regional earthquakes}.{\BBCQ}
\newblock
\APACjournalVolNumPages{Geophysical Research Letters}{37}{}{L16309}.
\PrintBackRefs{\CurrentBib}

\bibitem [\protect \citeauthoryear {%
Han%
, Vidale%
, Houston%
, Chao%
\BCBL {}\ \BBA {} Obara%
}{%
Han%
\ \protect \BOthers {.}}{%
{\protect \APACyear {2014}}%
}]{%
HAN_2014}
\APACinsertmetastar {%
HAN_2014}%
\begin{APACrefauthors}%
Han, J.%
, Vidale, J\BPBI E.%
, Houston, H.%
, Chao, K.%
\BCBL {}\ \BBA {} Obara, K.%
\end{APACrefauthors}%
\unskip\
\newblock
\APACrefYearMonthDay{2014}{}{}.
\newblock
{\BBOQ}\APACrefatitle {Triggering of tremor and inferred slow slip by small
  earthquakes at the {Nankai} subduction zone in southwest {Japan}} {Triggering
  of tremor and inferred slow slip by small earthquakes at the {Nankai}
  subduction zone in southwest {Japan}}.{\BBCQ}
\newblock
\APACjournalVolNumPages{Geophysical Research Letters}{41}{}{8053-8060}.
\PrintBackRefs{\CurrentBib}

\bibitem [\protect \citeauthoryear {%
Houston%
}{%
Houston%
}{%
{\protect \APACyear {2015}}%
}]{%
HOU_2015}
\APACinsertmetastar {%
HOU_2015}%
\begin{APACrefauthors}%
Houston, H.%
\end{APACrefauthors}%
\unskip\
\newblock
\APACrefYearMonthDay{2015}{}{}.
\newblock
{\BBOQ}\APACrefatitle {Low friction and fault weakening revealed by rising
  sensitivity of tremor to tidal stress} {Low friction and fault weakening
  revealed by rising sensitivity of tremor to tidal stress}.{\BBCQ}
\newblock
\APACjournalVolNumPages{Nature Geoscience}{8}{5}{409-415}.
\PrintBackRefs{\CurrentBib}

\bibitem [\protect \citeauthoryear {%
Houston%
, Delbridge%
, Wech%
\BCBL {}\ \BBA {} Creager%
}{%
Houston%
\ \protect \BOthers {.}}{%
{\protect \APACyear {2011}}%
}]{%
HOU_2011}
\APACinsertmetastar {%
HOU_2011}%
\begin{APACrefauthors}%
Houston, H.%
, Delbridge, B\BPBI G.%
, Wech, A\BPBI G.%
\BCBL {}\ \BBA {} Creager, K\BPBI C.%
\end{APACrefauthors}%
\unskip\
\newblock
\APACrefYearMonthDay{2011}{}{}.
\newblock
{\BBOQ}\APACrefatitle {Rapid tremor reversals in {Cascadia} generated by a
  weakened plate interface} {Rapid tremor reversals in {Cascadia} generated by
  a weakened plate interface}.{\BBCQ}
\newblock
\APACjournalVolNumPages{Nature Geoscience}{4}{}{404-409}.
\PrintBackRefs{\CurrentBib}

\bibitem [\protect \citeauthoryear {%
Ide%
, Shelly%
\BCBL {}\ \BBA {} Beroza%
}{%
Ide%
\ \protect \BOthers {.}}{%
{\protect \APACyear {2007}}%
}]{%
IDE_2007_GRL}
\APACinsertmetastar {%
IDE_2007_GRL}%
\begin{APACrefauthors}%
Ide, S.%
, Shelly, D\BPBI R.%
\BCBL {}\ \BBA {} Beroza, G\BPBI C.%
\end{APACrefauthors}%
\unskip\
\newblock
\APACrefYearMonthDay{2007}{}{}.
\newblock
{\BBOQ}\APACrefatitle {Mechanism of deep low frequency earthquakes: Further
  evidence that deep non-volcanic tremor is generated by shear slip on the
  plate interface} {Mechanism of deep low frequency earthquakes: Further
  evidence that deep non-volcanic tremor is generated by shear slip on the
  plate interface}.{\BBCQ}
\newblock
\APACjournalVolNumPages{Geophysical Research Letters}{34}{}{L03308}.
\PrintBackRefs{\CurrentBib}

\bibitem [\protect \citeauthoryear {%
Kato%
\ \BBA {} Nakagawa%
}{%
Kato%
\ \BBA {} Nakagawa%
}{%
{\protect \APACyear {2020}}%
}]{%
KAT_2020}
\APACinsertmetastar {%
KAT_2020}%
\begin{APACrefauthors}%
Kato, A.%
\BCBT {}\ \BBA {} Nakagawa, S.%
\end{APACrefauthors}%
\unskip\
\newblock
\APACrefYearMonthDay{2020}{}{}.
\newblock
{\BBOQ}\APACrefatitle {Detection of deep low-frequency earthquakes in the
  {Nankai} subduction zone over 11 years using a matched filter technique}
  {Detection of deep low-frequency earthquakes in the {Nankai} subduction zone
  over 11 years using a matched filter technique}.{\BBCQ}
\newblock
\APACjournalVolNumPages{Earth, Planets and Space}{72}{}{128}.
\PrintBackRefs{\CurrentBib}

\bibitem [\protect \citeauthoryear {%
Katsumata%
\ \BBA {} Kamaya%
}{%
Katsumata%
\ \BBA {} Kamaya%
}{%
{\protect \APACyear {2003}}%
}]{%
KAT_2003}
\APACinsertmetastar {%
KAT_2003}%
\begin{APACrefauthors}%
Katsumata, A.%
\BCBT {}\ \BBA {} Kamaya, N.%
\end{APACrefauthors}%
\unskip\
\newblock
\APACrefYearMonthDay{2003}{}{}.
\newblock
{\BBOQ}\APACrefatitle {Low-frequency continuous tremor around the {Moho}
  discontinuity away from volcanoes in the southwest {Japan}} {Low-frequency
  continuous tremor around the {Moho} discontinuity away from volcanoes in the
  southwest {Japan}}.{\BBCQ}
\newblock
\APACjournalVolNumPages{Geophysical Research Letters}{30}{1}{1020}.
\PrintBackRefs{\CurrentBib}

\bibitem [\protect \citeauthoryear {%
{Levander, A.}%
}{%
{Levander, A.}%
}{%
{\protect \APACyear {2007}}%
}]{%
https://doi.org/10.7914/sn/xq_2007}
\APACinsertmetastar {%
https://doi.org/10.7914/sn/xq_2007}%
\begin{APACrefauthors}%
{Levander, A.}%
\end{APACrefauthors}%
\unskip\
\newblock
\APACrefYearMonthDay{2007}{}{}.
\newblock
\APACrefbtitle {Seismic and Geodetic Investigations of Mendocino Triple
  Junction Dynamics.} {Seismic and geodetic investigations of mendocino triple
  junction dynamics.}
\newblock
\APACaddressPublisher{}{International Federation of Digital Seismograph
  Networks}.
\newblock
\begin{APACrefURL} \url{https://www.fdsn.org/networks/detail/XQ_2007/}
  \end{APACrefURL}
\newblock
\begin{APACrefDOI} \doi{10.7914/SN/XQ_2007} \end{APACrefDOI}
\PrintBackRefs{\CurrentBib}

\bibitem [\protect \citeauthoryear {%
McCrory%
, Blair%
, Oppenheimer%
\BCBL {}\ \BBA {} Walter%
}{%
McCrory%
\ \protect \BOthers {.}}{%
{\protect \APACyear {2006}}%
}]{%
MCC_2006}
\APACinsertmetastar {%
MCC_2006}%
\begin{APACrefauthors}%
McCrory, P\BPBI A.%
, Blair, J\BPBI L.%
, Oppenheimer, D\BPBI H.%
\BCBL {}\ \BBA {} Walter, S\BPBI R.%
\end{APACrefauthors}%
\unskip\
\newblock
\APACrefYearMonthDay{2006}{}{}.
\newblock
\APACrefbtitle {Depth to the {Juan} de {Fuca} slab beneath the {Cascadia}
  subduction margin - A {3-D} model sorting earthquakes} {Depth to the {Juan}
  de {Fuca} slab beneath the {Cascadia} subduction margin - a {3-D} model
  sorting earthquakes}\ \APACbVolEdTR{}{\BTR{}\ \BNUM\ Data Series 91}.
\newblock
\APACaddressInstitution{}{U.S. Geological Survey}.
\PrintBackRefs{\CurrentBib}

\bibitem [\protect \citeauthoryear {%
Miyazawa%
, Brodsky%
\BCBL {}\ \BBA {} Mori%
}{%
Miyazawa%
\ \protect \BOthers {.}}{%
{\protect \APACyear {2008}}%
}]{%
MIY_2008}
\APACinsertmetastar {%
MIY_2008}%
\begin{APACrefauthors}%
Miyazawa, M.%
, Brodsky, E\BPBI E.%
\BCBL {}\ \BBA {} Mori, J.%
\end{APACrefauthors}%
\unskip\
\newblock
\APACrefYearMonthDay{2008}{}{}.
\newblock
{\BBOQ}\APACrefatitle {Learning from dynamic triggering of low-frequency tremor
  in subduction zones} {Learning from dynamic triggering of low-frequency
  tremor in subduction zones}.{\BBCQ}
\newblock
\APACjournalVolNumPages{Earth, Planets and Space}{60}{}{e17-e20}.
\PrintBackRefs{\CurrentBib}

\bibitem [\protect \citeauthoryear {%
Nakamura%
}{%
Nakamura%
}{%
{\protect \APACyear {2017}}%
}]{%
NAK_2017}
\APACinsertmetastar {%
NAK_2017}%
\begin{APACrefauthors}%
Nakamura, M.%
\end{APACrefauthors}%
\unskip\
\newblock
\APACrefYearMonthDay{2017}{}{}.
\newblock
{\BBOQ}\APACrefatitle {Distribution of low-frequency earthquakes accompanying
  the very low frequency earthquakes along the {Ryukyu} {Trench}} {Distribution
  of low-frequency earthquakes accompanying the very low frequency earthquakes
  along the {Ryukyu} {Trench}}.{\BBCQ}
\newblock
\APACjournalVolNumPages{Earth, Planets, and Space}{69}{1}{1-17}.
\PrintBackRefs{\CurrentBib}

\bibitem [\protect \citeauthoryear {%
Obara%
}{%
Obara%
}{%
{\protect \APACyear {2002}}%
}]{%
OBA_2002}
\APACinsertmetastar {%
OBA_2002}%
\begin{APACrefauthors}%
Obara, K.%
\end{APACrefauthors}%
\unskip\
\newblock
\APACrefYearMonthDay{2002}{}{}.
\newblock
{\BBOQ}\APACrefatitle {Nonvolcanic deep tremor associated with subduction in
  southwest {Japan}} {Nonvolcanic deep tremor associated with subduction in
  southwest {Japan}}.{\BBCQ}
\newblock
\APACjournalVolNumPages{Science}{296}{5573}{1679-1681}.
\PrintBackRefs{\CurrentBib}

\bibitem [\protect \citeauthoryear {%
Obara%
, Hirose%
, Yamamizu%
\BCBL {}\ \BBA {} Kasahara%
}{%
Obara%
\ \protect \BOthers {.}}{%
{\protect \APACyear {2004}}%
}]{%
OBA_2004}
\APACinsertmetastar {%
OBA_2004}%
\begin{APACrefauthors}%
Obara, K.%
, Hirose, H.%
, Yamamizu, F.%
\BCBL {}\ \BBA {} Kasahara, K.%
\end{APACrefauthors}%
\unskip\
\newblock
\APACrefYearMonthDay{2004}{}{}.
\newblock
{\BBOQ}\APACrefatitle {Episodic slow slip events accompanied by non-volcanic
  tremors in southwest {Japan} subduction zone} {Episodic slow slip events
  accompanied by non-volcanic tremors in southwest {Japan} subduction
  zone}.{\BBCQ}
\newblock
\APACjournalVolNumPages{Geophysical Research Letters}{31}{}{L23602}.
\PrintBackRefs{\CurrentBib}

\bibitem [\protect \citeauthoryear {%
Ohta%
\ \BBA {} Ide%
}{%
Ohta%
\ \BBA {} Ide%
}{%
{\protect \APACyear {2017}}%
}]{%
OHT_2017}
\APACinsertmetastar {%
OHT_2017}%
\begin{APACrefauthors}%
Ohta, K.%
\BCBT {}\ \BBA {} Ide, S.%
\end{APACrefauthors}%
\unskip\
\newblock
\APACrefYearMonthDay{2017}{}{}.
\newblock
{\BBOQ}\APACrefatitle {Resolving the detailed spatiotemporal slip evolution of
  deep tremor in {Western} {Japan}} {Resolving the detailed spatiotemporal slip
  evolution of deep tremor in {Western} {Japan}}.{\BBCQ}
\newblock
\APACjournalVolNumPages{Journal of Geophysical Research: Solid
  Earth}{122}{12}{10009-10036}.
\PrintBackRefs{\CurrentBib}

\bibitem [\protect \citeauthoryear {%
Peng%
, Vidale%
, Wech%
, Nadeau%
\BCBL {}\ \BBA {} Creager%
}{%
Peng%
\ \protect \BOthers {.}}{%
{\protect \APACyear {2009}}%
}]{%
PEN_2009}
\APACinsertmetastar {%
PEN_2009}%
\begin{APACrefauthors}%
Peng, Z.%
, Vidale, J\BPBI E.%
, Wech, A\BPBI G.%
, Nadeau, R\BPBI M.%
\BCBL {}\ \BBA {} Creager, K\BPBI C.%
\end{APACrefauthors}%
\unskip\
\newblock
\APACrefYearMonthDay{2009}{}{}.
\newblock
{\BBOQ}\APACrefatitle {Remote triggering of tremor along the {San} {Andreas}
  {Fault} in central {California}} {Remote triggering of tremor along the {San}
  {Andreas} {Fault} in central {California}}.{\BBCQ}
\newblock
\APACjournalVolNumPages{Journal of Geophysical Research}{114}{}{B00A06}.
\PrintBackRefs{\CurrentBib}

\bibitem [\protect \citeauthoryear {%
Plourde%
, Bostock%
, Audet%
\BCBL {}\ \BBA {} Thomas%
}{%
Plourde%
\ \protect \BOthers {.}}{%
{\protect \APACyear {2015}}%
}]{%
PLO_2015}
\APACinsertmetastar {%
PLO_2015}%
\begin{APACrefauthors}%
Plourde, A\BPBI P.%
, Bostock, M\BPBI G.%
, Audet, P.%
\BCBL {}\ \BBA {} Thomas, A\BPBI M.%
\end{APACrefauthors}%
\unskip\
\newblock
\APACrefYearMonthDay{2015}{}{}.
\newblock
{\BBOQ}\APACrefatitle {Low-frequency earthquakes at the southern {Cascadia}
  margin} {Low-frequency earthquakes at the southern {Cascadia} margin}.{\BBCQ}
\newblock
\APACjournalVolNumPages{Geophysical Research Letters}{42}{}{4849-4855}.
\PrintBackRefs{\CurrentBib}

\bibitem [\protect \citeauthoryear {%
Rogers%
\ \BBA {} Dragert%
}{%
Rogers%
\ \BBA {} Dragert%
}{%
{\protect \APACyear {2003}}%
}]{%
ROG_2003}
\APACinsertmetastar {%
ROG_2003}%
\begin{APACrefauthors}%
Rogers, G.%
\BCBT {}\ \BBA {} Dragert, H.%
\end{APACrefauthors}%
\unskip\
\newblock
\APACrefYearMonthDay{2003}{}{}.
\newblock
{\BBOQ}\APACrefatitle {Tremor and slip on the {Cascadia} subduction zone: The
  chatter of silent slip} {Tremor and slip on the {Cascadia} subduction zone:
  The chatter of silent slip}.{\BBCQ}
\newblock
\APACjournalVolNumPages{Science}{300}{5627}{1942-1943}.
\PrintBackRefs{\CurrentBib}

\bibitem [\protect \citeauthoryear {%
Royer%
\ \BBA {} Bostock%
}{%
Royer%
\ \BBA {} Bostock%
}{%
{\protect \APACyear {2014}}%
}]{%
ROY_2014}
\APACinsertmetastar {%
ROY_2014}%
\begin{APACrefauthors}%
Royer, A\BPBI A.%
\BCBT {}\ \BBA {} Bostock, M\BPBI G.%
\end{APACrefauthors}%
\unskip\
\newblock
\APACrefYearMonthDay{2014}{}{}.
\newblock
{\BBOQ}\APACrefatitle {A comparative study of low frequency earthquake
  templates in northern {Cascadia}} {A comparative study of low frequency
  earthquake templates in northern {Cascadia}}.{\BBCQ}
\newblock
\APACjournalVolNumPages{Earth and Planetary Science Letters}{402}{}{247-256}.
\PrintBackRefs{\CurrentBib}

\bibitem [\protect \citeauthoryear {%
Royer%
, Thomas%
\BCBL {}\ \BBA {} Bostock%
}{%
Royer%
\ \protect \BOthers {.}}{%
{\protect \APACyear {2015}}%
}]{%
ROY_2015}
\APACinsertmetastar {%
ROY_2015}%
\begin{APACrefauthors}%
Royer, A\BPBI A.%
, Thomas, A\BPBI M.%
\BCBL {}\ \BBA {} Bostock, M\BPBI G.%
\end{APACrefauthors}%
\unskip\
\newblock
\APACrefYearMonthDay{2015}{}{}.
\newblock
{\BBOQ}\APACrefatitle {Tidal modulation and triggering of low-frequency
  earthquakes in northern {Cascadia}} {Tidal modulation and triggering of
  low-frequency earthquakes in northern {Cascadia}}.{\BBCQ}
\newblock
\APACjournalVolNumPages{Journal of Geophysical Research Solid
  Earth}{120}{}{384-405}.
\PrintBackRefs{\CurrentBib}

\bibitem [\protect \citeauthoryear {%
Rubinstein%
\ \protect \BOthers {.}}{%
Rubinstein%
\ \protect \BOthers {.}}{%
{\protect \APACyear {2009}}%
}]{%
RUB_2009}
\APACinsertmetastar {%
RUB_2009}%
\begin{APACrefauthors}%
Rubinstein, J\BPBI L.%
, Gomberg, J.%
, Vidale, J\BPBI E.%
, Wech, A\BPBI G.%
, Kao, H.%
, Creager, K\BPBI C.%
\BCBL {}\ \BBA {} Rogers, G.%
\end{APACrefauthors}%
\unskip\
\newblock
\APACrefYearMonthDay{2009}{}{}.
\newblock
{\BBOQ}\APACrefatitle {Seismic wave triggering of nonvolcanic tremor, episodic
  tremor and slip, and earthquakes on {Vancouver} {Island}} {Seismic wave
  triggering of nonvolcanic tremor, episodic tremor and slip, and earthquakes
  on {Vancouver} {Island}}.{\BBCQ}
\newblock
\APACjournalVolNumPages{Journal of Geophysical Research Solid
  Earth}{114}{}{B00A01}.
\PrintBackRefs{\CurrentBib}

\bibitem [\protect \citeauthoryear {%
Shelly%
}{%
Shelly%
}{%
{\protect \APACyear {2017}}%
}]{%
SHE_2017}
\APACinsertmetastar {%
SHE_2017}%
\begin{APACrefauthors}%
Shelly, D\BPBI R.%
\end{APACrefauthors}%
\unskip\
\newblock
\APACrefYearMonthDay{2017}{}{}.
\newblock
{\BBOQ}\APACrefatitle {A 15 year catalog of more than 1 million low-frequency
  earthquakes: Tracking tremor and slip along the deep {San} {Andreas} {Fault}}
  {A 15 year catalog of more than 1 million low-frequency earthquakes: Tracking
  tremor and slip along the deep {San} {Andreas} {Fault}}.{\BBCQ}
\newblock
\APACjournalVolNumPages{Journal of Geophysical Research: Solid
  Earth}{122}{}{3739-3753}.
\PrintBackRefs{\CurrentBib}

\bibitem [\protect \citeauthoryear {%
Shelly%
, Beroza%
\BCBL {}\ \BBA {} Ide%
}{%
Shelly%
\ \protect \BOthers {.}}{%
{\protect \APACyear {2007a}}%
}]{%
SHE_2007_nature}
\APACinsertmetastar {%
SHE_2007_nature}%
\begin{APACrefauthors}%
Shelly, D\BPBI R.%
, Beroza, G\BPBI C.%
\BCBL {}\ \BBA {} Ide, S.%
\end{APACrefauthors}%
\unskip\
\newblock
\APACrefYearMonthDay{2007a}{}{}.
\newblock
{\BBOQ}\APACrefatitle {Non-volcanic tremor and low-frequency earthquake swarms}
  {Non-volcanic tremor and low-frequency earthquake swarms}.{\BBCQ}
\newblock
\APACjournalVolNumPages{Nature}{446}{}{305-307}.
\PrintBackRefs{\CurrentBib}

\bibitem [\protect \citeauthoryear {%
Shelly%
, Beroza%
\BCBL {}\ \BBA {} Ide%
}{%
Shelly%
\ \protect \BOthers {.}}{%
{\protect \APACyear {2007b}}%
}]{%
SHE_2007_G3}
\APACinsertmetastar {%
SHE_2007_G3}%
\begin{APACrefauthors}%
Shelly, D\BPBI R.%
, Beroza, G\BPBI C.%
\BCBL {}\ \BBA {} Ide, S.%
\end{APACrefauthors}%
\unskip\
\newblock
\APACrefYearMonthDay{2007b}{}{}.
\newblock
{\BBOQ}\APACrefatitle {Complex evolution of transient slip derived from precise
  tremor locations in western {Shikoku}, {Japan}} {Complex evolution of
  transient slip derived from precise tremor locations in western {Shikoku},
  {Japan}}.{\BBCQ}
\newblock
\APACjournalVolNumPages{Geochemistry, Geophysics, Geosystems}{8}{}{Q10014}.
\PrintBackRefs{\CurrentBib}

\bibitem [\protect \citeauthoryear {%
Shelly%
, Beroza%
, Ide%
\BCBL {}\ \BBA {} Nakamula%
}{%
Shelly%
\ \protect \BOthers {.}}{%
{\protect \APACyear {2006}}%
}]{%
SHE_2006}
\APACinsertmetastar {%
SHE_2006}%
\begin{APACrefauthors}%
Shelly, D\BPBI R.%
, Beroza, G\BPBI C.%
, Ide, S.%
\BCBL {}\ \BBA {} Nakamula, S.%
\end{APACrefauthors}%
\unskip\
\newblock
\APACrefYearMonthDay{2006}{}{}.
\newblock
{\BBOQ}\APACrefatitle {Low-frequency earthquakes in {Shikoku}, {Japan}, and
  their relationship to episodic tremor and slip} {Low-frequency earthquakes in
  {Shikoku}, {Japan}, and their relationship to episodic tremor and
  slip}.{\BBCQ}
\newblock
\APACjournalVolNumPages{Nature}{442}{}{188-192}.
\PrintBackRefs{\CurrentBib}

\bibitem [\protect \citeauthoryear {%
Sweet%
, Creager%
, Houston%
\BCBL {}\ \BBA {} Chestler%
}{%
Sweet%
\ \protect \BOthers {.}}{%
{\protect \APACyear {2019}}%
}]{%
SWE_2019}
\APACinsertmetastar {%
SWE_2019}%
\begin{APACrefauthors}%
Sweet, J\BPBI R.%
, Creager, K\BPBI C.%
, Houston, H.%
\BCBL {}\ \BBA {} Chestler, S\BPBI R.%
\end{APACrefauthors}%
\unskip\
\newblock
\APACrefYearMonthDay{2019}{}{}.
\newblock
{\BBOQ}\APACrefatitle {Variations in {Cascadia} low-frequency earthquake
  behavior with downdip distance} {Variations in {Cascadia} low-frequency
  earthquake behavior with downdip distance}.{\BBCQ}
\newblock
\APACjournalVolNumPages{Geochemistry, Geophysics, Geosystems}{20}{}{1202-1217}.
\PrintBackRefs{\CurrentBib}

\bibitem [\protect \citeauthoryear {%
Thomas%
, B\"urgmann%
, Shelly%
, Beeler%
\BCBL {}\ \BBA {} Rudolph%
}{%
Thomas%
\ \protect \BOthers {.}}{%
{\protect \APACyear {2012}}%
}]{%
THO_2012}
\APACinsertmetastar {%
THO_2012}%
\begin{APACrefauthors}%
Thomas, A\BPBI M.%
, B\"urgmann, R.%
, Shelly, D\BPBI R.%
, Beeler, N\BPBI M.%
\BCBL {}\ \BBA {} Rudolph, M\BPBI L.%
\end{APACrefauthors}%
\unskip\
\newblock
\APACrefYearMonthDay{2012}{}{}.
\newblock
{\BBOQ}\APACrefatitle {Tidal triggering of low frequency earthquakes near
  {Parkfield}, {California}: Implications for fault mechanics within the
  brittle-ductile transition} {Tidal triggering of low frequency earthquakes
  near {Parkfield}, {California}: Implications for fault mechanics within the
  brittle-ductile transition}.{\BBCQ}
\newblock
\APACjournalVolNumPages{Journal of Geophysical Research}{117}{}{B05301}.
\PrintBackRefs{\CurrentBib}

\bibitem [\protect \citeauthoryear {%
Wech%
\ \BBA {} Creager%
}{%
Wech%
\ \BBA {} Creager%
}{%
{\protect \APACyear {2011}}%
}]{%
WEC_2011}
\APACinsertmetastar {%
WEC_2011}%
\begin{APACrefauthors}%
Wech, A\BPBI G.%
\BCBT {}\ \BBA {} Creager, K\BPBI C.%
\end{APACrefauthors}%
\unskip\
\newblock
\APACrefYearMonthDay{2011}{}{}.
\newblock
{\BBOQ}\APACrefatitle {A continuum of stress, strength and slip in the
  {Cascadia} subduction zone} {A continuum of stress, strength and slip in the
  {Cascadia} subduction zone}.{\BBCQ}
\newblock
\APACjournalVolNumPages{Nature Geoscience}{4}{}{624-628}.
\PrintBackRefs{\CurrentBib}

\bibitem [\protect \citeauthoryear {%
Wessel%
\ \BBA {} Smith%
}{%
Wessel%
\ \BBA {} Smith%
}{%
{\protect \APACyear {1991}}%
}]{%
WES_1991}
\APACinsertmetastar {%
WES_1991}%
\begin{APACrefauthors}%
Wessel, P.%
\BCBT {}\ \BBA {} Smith, W\BPBI H\BPBI F.%
\end{APACrefauthors}%
\unskip\
\newblock
\APACrefYearMonthDay{1991}{}{}.
\newblock
{\BBOQ}\APACrefatitle {Free software helps map and display data} {Free software
  helps map and display data}.{\BBCQ}
\newblock
\APACjournalVolNumPages{EOS Trans. AGU}{72}{}{441}.
\PrintBackRefs{\CurrentBib}

\end{thebibliography}

%
%
%
%
%

\end{document}